# High Performance Computing in Medical Image Analysis – HuSSaR*

László Kovács[1], Roland Kovács[2] and András Hajdu[3]

*Abstract*— In our former works we have made serious efforts to improve the performance of medical image analysis methods with using ensemble-based systems. In this paper, we present a novel hardware-based solution for the efficient adoption of our complex, fusion-based approaches for real-time applications. Even though most of the image processing problems and the increasing amount of data have high-performance computing (HPC) demand, there is still a lack of corresponding dedicated HPC solutions for several medical tasks. To widen this bottleneck we have developed a Hybrid Small Size high performance computing Resource (abbreviated by HuSSaR) which efficiently alloys CPU and GPU technologies, mobile and has an own cooling system to support easy mobility and wide applicability. Besides a proper technical description, we include several practical examples from the clinical data processing domain in this work. For more details see also: `https://arato.inf.unideb.hu/kovacs.laszlo/research_hybridmicrohpc.html`

## I. INTRODUCTION

Information technology has developed significantly, especially in the last decades. New hardware-based achievements establish never seen capabilities for more precise detection and personalized treatment in healthcare. General-purpose computing on graphics processing units (GPGPU)[1], the Intel Many Integrated Core Architecture (MIC)[2], the field-programmable gate array (FPGA)[3] and more recently the Neural Processing Unit (NPU) architectures are the examples for the hardware related progress. Several papers have been published about performance comparisons in different applications to determine the superiority of the Processing Units (PU)[4]. Though comparative studies [5] suggest it, the performance superiority of this platform is not really obvious in several cases. Namely, though GPUs are capable of running faster for instance deep learning algorithms than other current PUs, it should be noticed that GPUs are specialized vector processors on which convolutional operations are very efficient. PUs can be taken advantage of using fusion-based approaches with higher computing efficiency instead of considering competitive individual algorithms. Although the area of medical image analysis is full of high-performance computing problems, including machine learning, deep learning, and ensemble-based systems, access to efficient, corresponding HPC systems for the daily routine is limited. Thus, solutions like the proposed Hybrid Small Size high performance computing Resource (HuSSaR) can be a precious tool.

*Hybrid Small Size HPC Resource.
[1]Email: `kovacs.laszlo@inf.unideb.hu`
[2]Email: `kovirolig@gmail.com`
[3]Email: `hajdu.andras@inf.unideb.hu`
[1,2,3]Faculty of Informatics, Computer Graphic and Image Processing Dept., University of Debrecen, 26 Kassai str, 4028 Debrecen

## II. THE HUSSAR MACHINE

Our main aim was to build a hybrid small size mobile high-performance computer which could be placed in an operation room or place on a truck for mobile screening to support different detection or treatment-related computational issues [6]. While the number of processor units (PU) is limited on the mainboards, the co-processor cards are the only options to be built a high PU density machine in a regular size case. When there are no concrete algorithm predefined, it is hard to tell in advance which hardware technology of PU should be followed. It is proved that the aggregation of different solutions can solve a complex problem more efficiently compared with any individual approach. HPC hardware components generate a high amount of heat, thus, they are designed to work in an air-conditioned server room. The maximum working temperature of these PUs is around 90°C. The HPC components are overheated during heavy loaded computations at common room temperature (20-25°C). To achieve our objective, we have constructed a high PU density machine equipped with combined liquid/airflow cooling, which is sufficiently efficient for HPC demands. Moreover, it prevents overheating, performance drop and noise intolerability caused by fans.

### A. Built-up

Although the field of hybrid computing is a new domain, there are several attempts to exploit PU architectures (CPU, GPU, MIC, FPGA) simultaneously. To guarantee full usability of the CPUs, HuSSaR has an additional co-processor for each CPU. The number of the co-processors could be determined based on the required bandwidth and connection type between the CPU and the co-processors, as well. We have considered a dual processor motherboard with two different types (GPGPU and MIC) of co-processor cards for each CPU to support hybrid computing on PCI-E sockets. The prototype was designed carefully to establish direct wired connections between the CPUs and their co-processors. This architecture supports the optimal assignment of the specific subtasks to the necessary co-processor **??**. Moreover, all the PUs can work as a separate computing node if needed, but can also form a CPU-GPGPU-MIC set to enhance performance ([5], [11],[12]).

*1) Cooling:* The maximum operating temperatures of PUs is about 90°C, which limit is easily reached, if e.g. a strong GPU card is inserted into a common PC. In such situations, the driver usually decreases performance. In a HPC environment, this problem is avoided with maintaining constant temperature via air-conditioning, so that the significant part of the heat is conducted by a liquid-based cooling system.

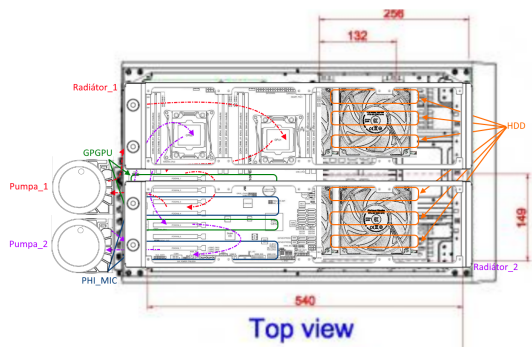

Fig. 1: Construction of HuSSaR, where red and purple arrows show the flow direction of the coolant.

Liquid(water)-based cooling is popular also for regular PCs primarily for optical tunning purposes. It might seem to be a costly hobby to build a customized, water-cooled PC only for optical tunning reasons, combining it with air-based ventilation we can reach HPC-level cooling. To follow the idea of the combination of PUs, we have created one cooling cycle for each CPU-GPGPU-MIC arrangement. Since our prototype includes two such PU sets, two separated parallel liquid cooling circuits have been considered. Every cooling circuit has a main cooling station (radiator) and optionally secondary one for each co-processor. Since in a cooling circuit the PUs are cooled sequentially, without additional cooling points the heat generated by the PUs may affect the cooling performance of the next PU in the circuit (Figure 1).

As the liquid circuit has the capability to remove the main part of the heat, directed airflow is sufficient to provide a stable cooling for the other components. This hybrid cooling is similar to large HPC cooling systems. For efficient cooling, the fans of the cooling points suck in the warm air from the inner region of the case and blow it out. This airflow is directed by considering that hot air has lower density, thus, fresh air can pass into the case mostly at the bottom, while the hot air leave the case at the top. To support this airflow, the hard disc drives are placed vertically, and there are fixed fans at the bottom of the case. A reservoir is also integrated with a relief valve and a high-performance pump to hold and move the coolant through the cooling circuits.

The speed of the pumps, fans are controlled with Pulse-width modulation (PWM) signals gathered from the mainboard and spread by a PWM hub controller. It is also possible to connect the cooling circuit to external cooling points in case of further cooling is needed. During our stress tests under full load of computations, the temperatures of the CPUs were around 40°C, GPUs 50°C, and Xeon Phi CPUs 60°C in a closed office with room temperature 24°C. The rotation speed of the pumps and fans were approximately 800RPM (possible maximum 3500RMP).

*2) Power:* The HuSSaR is designed to have two power supply units (PSU). The main one has the responsibility to supply the electricity for the main components (e.g. PUs, motherboard), while the secondary for the cooling circuits (pumps, fans). The main PSU switches on automatically, while the secondary when HuSSaR is turned on. After the system is halted, the secondary PSU remains under power to cool down the components. The minimum required power of the PSU depends mostly on the sum of the required power of the PUs. In our prototype, the main PSU has the capacity 1700W, while the secondary one 450W.

## III. APPLICATIONS WITH OPTIMAL TASK SCHEDULING

The idea of hybrid applications is twofold. First, the types of HPC environments based on different architectures increase according to the rapidly developing software and hardware technologies. Thus, attempting optimal usage of the available tools is a natural motivation. Second, complex systems that aggregate different approaches could outperform single solution based ones [11]. However, such fusion-based approaches have large computational needs with including more algorithms or complex work-flows. As for modeling this approach, we can consider a set of components to be executed with their corresponding computational needs. If we have limited computational resources and an executional time constraint for execution, the optimal selection of a subset of the components can be solved as a non-linear Knapsack problem. For more details on the theoretical results on this topic, see [12]. As a final outcome of this step, we have the components that needs computational resources to be assigned to them.

To control the access to the computing resources, HuSSaR considers state-of-the-art schedulers to support distributed processing e.g. for a fusion-based application consisting of many algorithms. Technically, the system has master and computing nodes, where the master receives and schedules the computing task via waiting queues. The free computing node (PU resources) can pick and execute the jobs from the queue. Since there is no shared memory access, this technique is used mainly for running the jobs with different inputs. If there is a need to increase the performance of a single HuSSaR machine, two or four of them can be piled up (see Figure 2b), where all the member HuSSaRs have from one to four CPUs equipped with the corresponding co-processors. The communication among the HuSSaR is realized via UTP network using scheduling or computing library software solutions. Overall, a HuSSaR configuration of maximum performance includes 4 motherboards, 16 CPUs, and 32 co-processor cards providing operation speed 350 TFlops.

Machine learning technologies nowadays have a leading role in the field of image processing. However, the application of deep learning-based methods is only one example. As a computational task, deep learning is just similar to other traditional e.g. detector algorithms or an ensemble of them for a dedicated image processing task. Thus, similarly to distributed deep learning, they can be efficiently executed

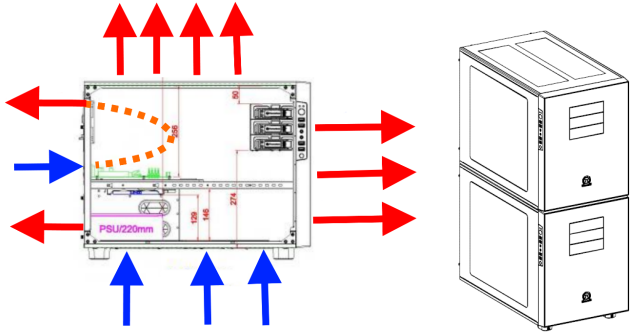

(a) Direction of airflow direction; blue/red arrows show the flow of fresh/hot air.
(b) Piling HuSSaRs.

Fig. 2: HuSSaR built-up.

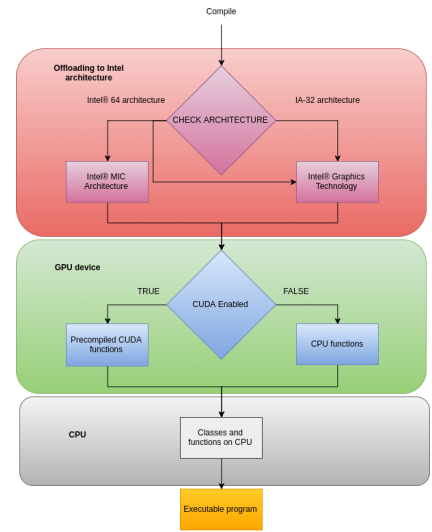

Fig. 3: Compiler architecture.

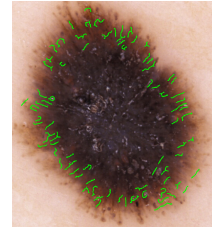

Fig. 4: Detected streaks in a mole.

on the HuSSaR system. Because of the brevity of paper, we demonstrate the operation of HuSSaR on a specific task corresponding to the detection of skin lesions in medical images [7], which procedure does not consider a learning step.

*1) Skin Lesion Detection:* Malignant melanoma is a cancerous lesion from the pigment-bearing basal layer of the skin. Melanoma is one of the deadliest form of skin cancer in the developed countries, though it can be treated by surgical excision if detected early. One of the visual features of melanoma is the presence of streaks. Streaks are linear extensions of a mole/pigmented lesion that are arranged in linear structures in the radial growth directions of the mole at its borders. This clinical definition can be translated to an abstract model, whose parameters can be determined by image processing techniques: 1) Streaks are three or more linear structures co-radially oriented at the border of the mole having thickness 1/3 of the minor axis of the lesion. 2) Streaks are darker than other mole regions in their neighborhood. 3) Streaks are shorter than the 1/3 of the minor axis of the lesion, and they should be longer than 1% of the major axis. 4) Streaks do not branch and have small curvature.

As an example to detect the streaks on melanoma images, we can use conventional image processing tools as follows. First, the lesion is segmented, rotated to align its major axis to be horizontal as it represents the growth direction of the mole. To detect dermoscopic structures, a Laplacian of Gaussian operator is applied to the image. A Gabor filter with orientation and tuned ridge frequency removes noise while preserves correct ridges and valleys. Since streaks appear in more radial directions, Gabor filters are considered in more directions (0°, 45°, 90°, 135°). A proper threshold function is used to get the main structures of the mole with 0 for valleys and 1 for ridges. As a final step, skeletonization is applied to get the streaks with a fast parallel algorithm [8] including a removal step for the too short/long skeleton segments, as well. A sample output can be observed in Figure 4.

To implement the above method to optimize performance, different parts of the corresponding codes run on different PUs to be computed at the same time. Namely, an image is processed using a CPU including Intel graphics cards (GFX) if exists, a GPU by Nvidia CUDA libraries, and an Intel MIC PU architecture, typically on Xeon Phi processors by Intel Parallel Libraries [9]. As Gaussian blur and minimum/maximum pixel intensity extraction run sufficiently fast on CPU, it is worthless uploading and downloading data to/from co-processor cards. However, Canny edge detector is fast on GPU, thus it is worth uploading/downloading. Figure 5 explains which part of the code has implemented on which architecture. The figure also shows how distributed parallelization is applied for each computing node; our prototype has two computing subnodes (two CPU/GPU/MIC sets).

*2) Compiler:* Because of the heterogeneous hardware setup there are no direct compiler options available. However, it is possible to combine the native compilers of the individual architectures, that is, Intel ICC and Nvidia Cuda. During compilation, there is a check to discover existing Intel and CUDA compatible architectures. The parameters and compiler commands are assembled in a makefile. The compilation procedure (see Figure 3) for co-processor architectures is executed before compiling the main functions of the code, which links the precompiled PU kernels directly.

*3) SWhybrid:* During the construction of HuSSaR, a study has been released [10] about a hybrid solution of optimal pairwise alignment of very long DNA sequences, and protein sequence database search. This work has also been shared on Github, and the hybrid implementation was successfully tested on HuSSaR without major problems.

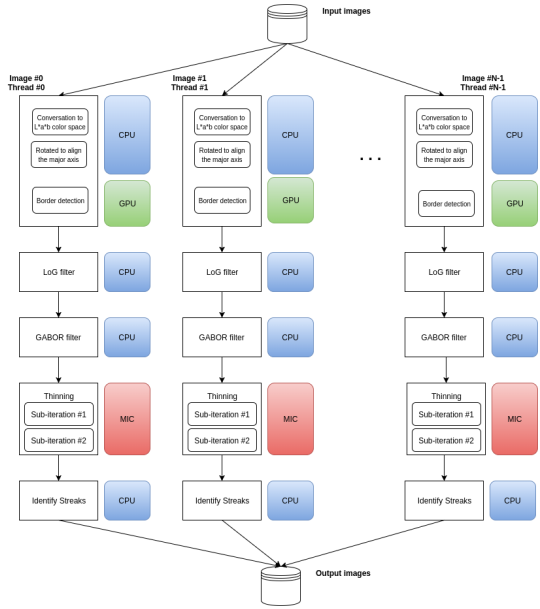

Fig. 5: Executional architecture.

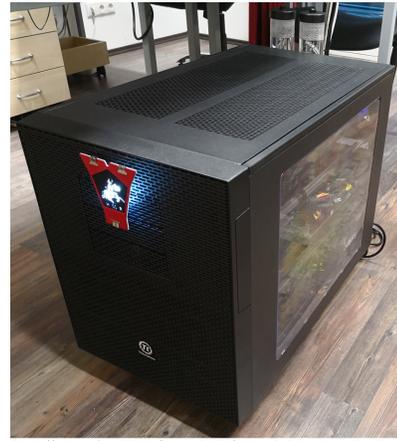

Fig. 6: HuSSaR machine

## IV. CONCLUSION

In this work, we have presented a Hybrid Small Size high-performance computing Resource (HuSSaR) dedicated mainly for digital image analysis. Our main motivation was to provide comfortable HPC access also for real-time applicability. To reach these aims, we have realized a hybrid air/liquid cooling. The performance of the cooling system makes HuSSaR mobilizable besides office usage. We have customized a double PSU system to match the power and cooling safety demands, where the primary and cooling system power are separated to keep up cooling after system halt. As a high-performance computing solution, we have applied a heterogeneous computing architecture to maximize hardware utilization capabilities. Compiling to the hybrid architecture is realized by the combination of the native compiler of PUs using kernel type precompile methodology. In our prototype of HuSSaR, we have considered Intel MIC and Nvidia GPU that are replaceable with another types of cards like FPGA or NPU. For distributed processing purposes, commonly applied schedulers are implemented to support e.g. ensemble-based systems. Machine learning algorithms including deep learning ones could be executed on HuSSaR, too. The optimal exploitation of the system was demonstrated on a skin lesion classification task. As future aims, we plan to optimize some of our former results on CT-based visualization [6] and creation of ensembles [12] to the HuSSaR (Figure 6) system.

## V. ACKNOWLEDGEMENT

The research was supported in part by Nvidia (Hardware grant), Surgi-Trade Kft., the projects GINOP-2.1.7-15-2016-01641 and EFOP-3.6.2-16-2017-00015 supported by the European Union and the State of Hungary, co-financed by the European Social Fund.